\documentclass[aps, prd, superscriptaddress, nofootinbib, preprintnumbers, showpacs]{revtex4}
\usepackage{graphicx}
\usepackage{amsmath}
\def\fsl#1{\setbox0=\hbox{$#1$}                 
   \dimen0=\wd0                                 
   \setbox1=\hbox{/} \dimen1=\wd1               
   \ifdim\dimen0>\dimen1                        
      \rlap{\hbox to \dimen0{\hfil/\hfil}}      
      #1                                        
   \else                                        
      \rlap{\hbox to \dimen1{\hfil$#1$\hfil}}   
      /                                         
   \fi}                                         %

\newcommand{\VEV}[1]{\langle #1 \rangle}

\preprint{UWO-TH-05/18}
\begin{document}
\title{Vortices in gauge models at finite density
with vector condensates}

\author{E.V. Gorbar}
  \email{egorbar@uwo.ca}
  \altaffiliation[On leave from ]{
       Bogolyubov Institute for Theoretical Physics,
       03143, Kiev, Ukraine}
\author{Junji Jia}
  \email{jjia5@uwo.ca}
\author{V.A. Miransky}
  \email{vmiransk@uwo.ca}
   \altaffiliation[On leave from ]{
       Bogolyubov Institute for Theoretical Physics,
       03143, Kiev, Ukraine}
\affiliation{
Department of Applied Mathematics, University of Western
Ontario, London, Ontario N6A 5B7, Canada
}

\date{\today}

\begin{abstract}
There exists a class of gauge models incorporating a finite density
of matter in which the Higgs mechanism is
provided by condensates of gauge (or gauge plus scalar) fields,
i.e., there are vector condensates in this case. We describe 
vortex solutions in the simplest model in this class,
the gauged $SU(2)\times U(1)_Y$
$\sigma$-model with the chemical potential for hypercharge $Y$,
in which the gauge symmetry is completely broken.
It is shown that there are three types of topologically stable vortices 
in the model, connected
either with photon field or hypercharge gauge field, or 
with both of them. Explicit
vortex solutions are numerically found and their energy per unit
length are calculated. The relevance of these solutions for the gluonic phase
in the dense two-flavor QCD is discussed. 
\end{abstract}


\maketitle

\section{Introduction}

Vortices are very special soliton-like excitations.
In condensed matter physics, vortex solutions were discovered by 
Abrikosov \cite{Abrikosov}: They play a crucial role in the dynamics of 
type II superconductors. In relativistic quantum
field theory, vortices were first considered  by Nielsen and Olesen
\cite{NO}. They can be important in cosmology, astroparticle
physics, and for the confinement dynamics in QCD-like theories
(for reviews, see Ref. \cite{VSh}). Vortex and vortex-like
solutions in relativistic models at finite density of quark matter
were studied in Ref. \cite{Zhit}.

The existence of vortex solutions
is usually connected with the Higgs mechanism in
abelian $U(1)$ gauge dynamics. The major degrees of freedom are
a complex scalar Higgs field and a $U(1)$ gauge field. In this 
paper, we will describe vortex solutions in a somewhat different
class of gauge models incorporating a finite density of matter.
The Higgs mechanism in these models is
provided by condensates of gauge (or gauge plus scalar) fields, 
i.e., there are vector condensates in this case. 
This class in particular
includes the $SU(2)_L\times U(1)_{Y}$ electroweak
theory in the presence of a superdense fermionic matter 
\cite{Linde,Ferrer}, (nonrenormalizable) models
including
massive vector bosons with a chemical potential for electric
charge \cite{Sannino}, and the gauged linear $SU(2)_L\times U(1)_Y$
$\sigma$-model (without fermions) with a chemical potential for
hypercharge $Y$ \cite{sigmamodel}. 
\footnote{Ungauged linear $SU(2)_L\times U(1)_{Y}$
$\sigma$-model with a chemical potential for
hypercharge \cite{MS}
is a toy model for the description of the dynamics of 
the kaon condensate in high density QCD \cite{BS}. 
In particular, it realizes the phenomenon with abnormal
number of Nambu-Goldstone (NG) bosons \cite{MS},
when spontaneous
breakdown of continuous symmetries leads to a lesser number of
NG bosons than that required by the Goldstone
theorem (for a recent discussion of this model, see
Ref. \cite{Brauner}).}

It is also noticeable that a recently revealed
gluonic phase in neutral two-flavor quark matter \cite{gluonic}
also relates to this class. In the gluonic phase, vector condensates of
gluons cure a chromomagnetic instability \cite{HS} in the two-flavor
superconducting (2SC)
solution and lead to spontaneous breakdown of the  
$SU(2)_c \times \tilde{U}(1)_{em}\times SO(3)_{\rm rot}$ symmetry
down to $SO(2)_{\rm rot}$. Here $SU(2)_c$ and $\tilde{U}(1)_{em}$ are
the color and electromagnetic gauge symmetries
in the 2SC medium, and $SO(3)_{\rm rot}$
is the rotational group (recall that in the 2SC solution the
color $SU(3)_c$ is broken down to $SU(2)_c$).
In other words, the gluonic phase describes an
anisotropic medium in which the color and electric superconductivities
coexist. Because it is naturally to
expect that cold quark matter may exist in the core of
compact stars (for reviews, see Ref. \cite{RW}), it would be of great 
interest to describe the dynamics of this phase in more detail,
in particular, to clarify whether vortex excitations exist there.

Since the dynamics of the gluonic phase
is very rich and complicated, as a first step,
it would be appropriate to use 
a simpler
model relating to the same universality class, i.e.,
with the same sample of the symmetry breaking. Fortunately,
such a model exists: It is
the gauged $\sigma$-model with the chemical potential for the hypercharge 
introduced and studied in Ref. \cite{sigmamodel}. 
Let us describe it in more detail.
Its Lagrangian density reads (we use the metric
$g^{\mu\nu}= \mbox{diag}(1, -1, -1, -1))$:
\begin{equation} {\cal L}=-\frac{1}{4}F^{(a)}_{\mu\nu}F^{\mu\nu(a)}-
\frac{1}{4}F^{(Y)}_{\mu\nu}F^{\mu\nu(Y)} +
[(D_{\nu}-i\mu\delta_{\nu0})\Phi]^{\dag}
(D^{\nu}-i\mu\delta^{\nu0})\Phi-m^2\Phi^{\dag}\Phi-\lambda(\Phi^{\dag}\Phi)^2,
\label{Lagrangian}
\end{equation}
where the covariant derivative $D_{\mu}=\partial_{\mu}-igA_{\mu}-
(ig^{\prime}/2)B_{\mu}$, $\Phi$ is a complex doublet field
$\Phi^T=(\varphi^+,\varphi_0)$, and the chemical
potential $\mu$ is provided by external conditions (to be specific, we take
$\mu >0$). Here
$A_{\mu}=A_{\mu}^a\tau^a/2$ are $SU(2)_L$ gauge fields ($\tau^a$ are 
three Pauli matrices) and the field strength
$F^{(a)}_{\mu\nu}=\partial_{\mu}A_{\nu}^{(a)}-
\partial_{\nu}A^{(a)}_{\mu} + g\epsilon^{abc}A^{(b)}_{\mu}A^{(c)}_{\nu}$.
$B_{\mu}$ is a $U_{Y}(1)$ 
gauge field with the field strength $F^{(Y)}_{\mu\nu}  =
\partial_{\mu}B_{\nu}-\partial_{\nu}B_{\mu}$. The hypercharge of the doublet
$\Phi$ equals +1. This model has the same
structure as the electroweak theory without fermions and with chemical
potential for hypercharge $Y$. 
Note that the terms with the chemical potential
are $SU(2)_L \times U(1)_Y$ (and not $SU(2)_{L}\times SU(2)_R$)
symmetric. This follows from the fact that
the hypercharge generator $Y$ is $Y = 2I^{3}_{R}$ where $I^{3}_{R}$
is the
third component of the right handed isospin generator. 
Henceforth
we will omit the subscripts $L$ and $R$, allowing various
interpretations of the $SU(2)$.

Because the chemical potential explicitly breaks
the Lorentz symmetry, the symmetry of the model is 
$SU(2) \times U(1)_Y \times SO(3)_{\rm rot}$.
As was shown in Ref. \cite{sigmamodel}, for sufficiently large
values of the chemical potential $\mu$, the condensates of
both the scalar doublet $\Phi$ {\it and} the gauge field $A_\mu$ occur. 
These condensates break the symmetry 
$SU(2) \times U(1)_Y \times SO(3)_{\rm rot}$
down to $SO(2)_{\rm rot}$. This is the same symmetry 
breaking pattern as in the gluonic phase of dense QCD 
\cite{gluonic}, while the gauged $\sigma$-model is much simpler:
It is renormalizable and for small coupling constants $g$,
$g^{\prime}$ and $\lambda$, the tree approximation is reliable there.

As will be 
shown below, 
topologically stable vortex solutions exist in this model indeed. Their
structure is much richer and more complicated than that in the
Abelian gauge model. In particular, there are different types of
vortices connected either with photon field or hypercharge
gauge field $B_{\mu}$, or with both of them. 

The paper is organized as follows. In Section \ref{two}, a general
analysis of vortex solutions in this model is realized. We discuss
a topological stability of these solutions and introduce three
types of vortices: Magnetic vortices, hypermagnetic vortices, and
hybrid ones. They are connected with photon field, hypercharge
field $B_{\mu}$, and both of them, respectively. In Section
\ref{three}, an explicit solution for magnetic vortices is
obtained. In Sections \ref{four} and \ref{five}, explicit
solutions for hypermagnetic vortices and hybrid ones, respectively, 
are derived. In Section \ref{six}, we summarize the main
results of the paper.

\section{Vortex solutions: General consideration}
\label{two}

The equations of motion in the gauged $\sigma$-model at finite density 
follow from Lagrangian (\ref{Lagrangian}):
\begin{equation}
-(D_{\nu}-i\mu\delta_{\nu0})(D^{\nu}-i\mu\delta^{\nu0})\Phi-m^2\Phi-
2\lambda(\Phi^{\dag}\Phi)\Phi = 0 \,,
\label{eqs-1-1}
\end{equation}
\begin{eqnarray}
\partial^{\mu}F^{(a)}_{\mu\nu}+ig\left[\Phi^{\dag}\frac{\tau^a}{2}
\left(\partial_{\nu}-i\frac{g'}{2}B_{\nu}\right)\Phi-
\left(\left(\partial_{\nu}-i\frac{g'}{2}B_{\nu}\right)\Phi\right)^{\dag}
\frac{\tau^a}{2}\Phi\right]+\nonumber\\
g\varepsilon^{abc}A^{\mu(b)}F^{(c)}_{\mu\nu}+\frac{g^2}{2}A_{\nu}^{(a)}
\Phi^{\dag}\Phi+2g\mu\delta_{\nu0}\Phi^{\dag}\frac{\tau^a}{2}\Phi = 0 \,,
\label{eqs-1-2}
\end{eqnarray}
\begin{equation}
\partial^{\mu}F^{(Y)}_{\mu\nu} +\frac{ig'}
{2}\Phi^{\dag}(D_{\nu}-i\mu\delta_{\nu0})
\Phi-\frac{ig'}{2}[(D_{\nu}-i\mu\delta_{\nu0})\Phi]^{\dag}\Phi = 0.
\label{eqs-1-3}
\end{equation}
As was shown in Ref. \cite{sigmamodel}, for sufficiently large values of
the chemical potential $\mu$, 
the vacuum solution of
(\ref{eqs-1-1})-(\ref{eqs-1-3}) is given by
\begin{equation}
W^{(-)}_z=(W^{(+)}_z)^* \equiv
C = \sqrt{\frac{\mu v_0}{\sqrt{2}g}-\frac{v_0^2}{4}},
\quad A^{(3)}_t = \frac{v_0}{\sqrt{2}}, \quad 
\Phi^T=(0,v_0),
\label{vacuum}
\end{equation}
where
\begin{equation}
v_0=\frac{\sqrt{(g^2+64\lambda)\mu^2-8(8\lambda-g^2)m^2}-3g\mu}{\sqrt{2}
(8\lambda-g^2)} \,,
\label{v0}
\end{equation}
$W^{(\mp)}_{\mu}=\frac{1}{\sqrt{2}}(A_{\mu}^{(1)} \pm iA_{\mu}^{(2)})$,
$\Phi^T=(\varphi^+,\varphi_0)$
and all other fields are equal to zero. 
\footnote {Here ``sufficiently large values of
$\mu$'' means the following: When $m^2 > 0$, $\mu$ should be larger
than the critical value $\mu_{cr} = m$, and for $m^2 < 0$,
$\mu$ should be larger
than $\mu_{cr} = g|m|/2\sqrt{\lambda}$ (the
critical value $g|m|/2\sqrt{\lambda}$ coincides with the mass
of $W$ boson in the vacuum theory with $\mu =0$ and $m^2 <0$).}  
It is clear that this solution implies that the initial symmetry
$SU(2) \times U(1)_Y \times SO(3)_{\rm rot}$ is spontaneously
broken down to $SO(2)_{\rm rot}$. In particular, the
electromagnetic $U(1)_{em}$, with electric charge 
$Q_{em} = I_3 + Y/2$,
is spontaneously broken by the condensate
of $W$ bosons, i.e., electric superconductivity takes place in
this medium.

Note that because the $U(1)_Y$ symmetry
is local, for nonzero chemical potential $\mu$
one should introduce a source term $B_0J_0$ in Lagrangian density
(\ref{Lagrangian}) in order to make the system neutral with respect to
hypercharge $Y$. This is necessary since otherwise in such a system 
thermodynamic equilibrium could not be
established. The value of the background hypercharge density $J_0$
(representing very heavy particles) is determined from the requirement that
$B_0=0$ is a solution of the equation of motion  
for $B_0$ (Gauss's law)
\cite{sigmamodel,Kapusta}. There exists an alternative description of
this dynamics in which a background hypercharge density $J_0$ is
considered as a free parameter and $\mu$ is taken to be zero. 
Then Gauss's law will define the vacuum expectation value 
$\VEV{B_0}$. It is not difficult to check that these two approaches
are equivalent if the chemical potential $\mu$ in the first 
approach is taken to be equal to the value
$\frac{g'}{2}\VEV{B_0}$ from
the second one. In this paper, following Refs. \cite{sigmamodel,Kapusta},
we use the first approach. 

As was shown in Ref. \cite{sigmamodel}, in accordance with  
the sample of spontaneous breakdown of the global symmetry, 
$SO(3)_{rot}\to SO(2)_{rot}$, there are two gapless 
NG bosons in this model. 
The other excitations are massive (the Higgs
mechanism). Because electric superconductivity
is realized in the model, it is naturally to expect that 
vortices may exist there. As will be shown in this paper, this is
indeed the case. 

Vortices are topologically nontrivial configurations with
fields approaching their vacuum values at spatial infinity \cite{VSh}.
Mathematically, vortices are topologically stable whenever the
first homotopy group $\pi_1(G_{int}/H_{int})$ of the
vacuum manifold $G_{int}/H_{int}$ is nontrivial (here $G_{int}$ is an 
internal symmetry group of the action in a model
and its subgroup $H_{int}$ is a symmetry group of the vacuum.) We
emphasize that this criterion of the topological stability is
sufficient but not necessary. In particular, as will be shown below,
magnetic vortices,
the most interesting vortex solutions in the present model,
are topologically stable due to another criterion. 

According to 
Eq. (\ref{vacuum}), two complex fields $W^{(-)}_z$ and
$\varphi_0$ have nonzero vacuum expectation values in the ground state of the
gauged $\sigma$-model at finite density. Therefore a priori there can be
different types of topologically stable vortex solutions: Both 
the field $W^{(-)}_z$ and the field $\varphi_0$ 
can wind around the spatial circle at infinity. I.e., in 
the cylindrical coordinates $(t, \rho, \phi, z)$, as
$\rho \to \infty$,
the asymptotics of these fields take the form
\begin{equation}
W^{(-)}_z \to Ce^{il\phi},\,\, \varphi_0 \to v_0e^{in\phi},
\label{gasympt}
\end{equation}
where $l$ and $n$ are integer. Let us turn to the analysis of a 
topological stability of 
solutions with these asymptotics. 

In this model, the internal
group $G_{int}$ is $SU(2)\times U(1)_Y$ and the symmetry group
$H_{int}$ of vacuum (\ref{vacuum}) is trivial. Therefore the first
homotopy group of
the vacuum manifold in the model is 
$\pi_1(G_{int}/H_{int}) = \pi_1(SU(2)\times U(1)_Y)$ =
$\pi_1(SU(2)) + \pi_1(U(1)_Y)$ = $\pi_1(U(1)_Y)$ = $Z$,
where $Z$ is the group of integer numbers (here we use
the fact that since the $SU(2)$ manifold is simply
connected, $\pi_1(SU(2))=0$). This seems to suggest
that there exist only topologically stable vortex solutions
connected with the hypercharge $U(1)_Y$. However, as will be
shown below, the situation in this model is more 
sophisticated and there exist also other topologically
stable vortices. 
The point is that while the scalar field
$\Phi$ is assigned to the fundamental representation of the
$SU(2)$ group, the gauge field $A_{\mu}^{(a)}$ is assigned to
the adjoint one. Because of that,
$SU(2)$ gauge transformations for $A_{\mu}^{(a)}$  reduce to $SO(3)$
ones. It is important that, unlike $SU(2)$, the manifold of
$SO(3)$ is not simply connected and $\pi_1(SO(3)) = Z_2$
with $Z_2 = 0, 1$.
As will become clear in a moment, 
this point is quite important in the analysis of
the topological stability of the solutions.

We begin
by considering the conventional case when $l = 0$ in
asymptotics (\ref{gasympt}):
\begin{equation}
W^{(-)}_z \to C,\,\, \varphi_0 \to v_0e^{in\phi}.
\label{hasympt}
\end{equation}
This configuration 
can be obtained from the vacuum configuration
(\ref{vacuum}) by using the $U(1)_Y$ gauge transformation
$e^{in\phi Y}$. It is clear that in this case the solutions 
with different $n$ are assigned to different topological classes
described by the homotopy group $\pi_1(U(1)_Y) = Z$. Therefore,
if such vortex solutions exist, they are topologically stable.
Due to the evident reasons, we will call them hypermagnetic vortices.
They possess hypermagnetic fluxes with a (hypermagnetic) winding
number $H=n$.

Let us now turn to a more interesting case of asymptotics (\ref{gasympt})
with $n=0$:
\begin{equation}
W^{(-)}_z \to Ce^{il\phi},\,\, \varphi_0 \to v_0.
\label{masympt}
\end{equation}
It is a rather unusual situation: The
phase $l\phi$ is now related to a vector field and not to a scalar one.
It is clear that configuration (\ref{masympt})
can be obtained from vacuum configuration
(\ref{vacuum}) by using the gauge transformation
\begin{equation}
e^{-il\phi Q_{em}} =  e^{-il\phi\tau^3/2}\,\,
e^{-il\phi Y/2},
\label{mtransf}
\end{equation}
where electric charge $Q_{em} = I_3 + Y/2$ is equal to -1 for
the field $W^{(-)}_z$ and zero for the scalar field $\varphi_0$.

Because for the vector fields the $SU(2)$ transformations are
reduced to the $SO(3)$ ones, it is important to distinguish two
cases: with $l$ being even and $l$ being odd. 
In the first case,
with $l = 2k$, as the angle $\phi$ running 
from 0 to
$2\pi$, the gauge transformation (\ref{mtransf}) defines 
closed loops both on the manifold of $SU(2)$ and 
that of $U(1)_{Y}$. 
Then,
because  
$\pi_1(SU(2)\times U(1)_Y) = \pi_1(U(1)_Y)$ =$Z$, we
conclude that these solutions are
topologically equivalent to the hypermagnetic vortex solutions
with the winding number $H=-k$ considered above.

The case with odd $l = 2k +1$ is more sophisticated. 
In this case, as $\phi$ changing from 0 to
$2\pi$, the gauge transformation (\ref{mtransf}) does
{\it not} define closed loops neither on the
$SU(2)$ manifold nor on the $U(1)_Y$ one. Therefore in this
case one cannot use the homotopic group $\pi_1(SU(2)\times U(1)_Y)$
for studying the topological stability of these solutions.
Let us show that for this purpose
one can use the homotopic group
$\pi_1(SO(3))$ instead. The reasons are the following.
Because the hypercharge of the gauge field $A_{\mu}^{(a)}$ 
is zero, the action of the gauge transformation (\ref{mtransf})
on it is given by the operator 
$e^{-i(2k+1)\phi I_{3}^{adj}/2}$, where
$I_{3}^{adj}$ is a $SU(2)$ generator in the adjoint representation.
As the angle $\phi$ changing from 0 to $2\pi$,
this transformation {\it does} define a closed loop on the
$SO(3)$ manifold, which coincides with the vacuum manifold  
of $A_{\mu}^{(a)}$. 
\footnote{These two manifolds coincide because vacuum configuration
(\ref{vacuum}) includes two linearly independent
isovectors, $A_{z}^{(1)}$, and 
$A_{t}^{(3)}$, and there is no subgroup in $SO(3)$ under which 
two linearly independent isovectors are invariant.}
Therefore, one can indeed
use the homotopic group $\pi_1(SO(3))=Z_2$
for the description of the topological stability of these
solutions. 

Now,
because odd values of $l$ correspond to the element 
+1 of $\pi_1(SO(3))=Z_2$, we conclude that this loop cannot be
unwinded and, therefore, these vortex solutions are topologically
stable. 
Due to the form of gauge transformation (\ref{mtransf}), 
we will call them magnetic
vortices. They possess magnetic fluxes with a 
(magnetic) winding number $M = -(2k + 1)$: The minus sign here
corresponds to a negative electric charge of $W^{-}$ bosons.
A detailed study of the 
topological structure  of the set of these solutions 
is beyond the scope of this paper. Here we will only argue
that the solutions with different values of $k$ are
topologically inequivalent. The argument is based on gauge
transformation (\ref{mtransf}). It implies that although for odd $l=2k+1$
the path defined by this transformation on the $U(1)_Y$ manifold is
not closed, it contains $k$ windings. Therefore the solutions
with different $k$ should be assigned to different topological classes,
i.e., they cannot be transformed from one into another by using a
continuous deformation. 

The last case is that with the ``hybrid'' asymptotics
\begin{equation}
W^{(-)}_z \to Ce^{il\phi},\,\, \varphi_0 \to v_0 e^{in\phi}
\label{hbasympt}
\end{equation}
with both $l$ and $n$ being nonzero.
In this case the gauge transformation (\ref{mtransf}) is
replaced by
\begin{equation}
e^{-il\phi Q_{em}}\,\, e^{in\phi Y} =  e^{-il\phi\tau^3/2}\,\,
e^{i(2n-l)\phi Y/2}.
\label{hbtransf}
\end{equation}
Using the same arguments as above, we conclude the following.
For even $l =2k$, these solutions are topologically
equivalent to the hypermagnetic vortices with 
fluxes corresponding to the winding number $H = n-k$. 
As to odd $l =2k + 1$, they yield new topological classes of
solutions. These solutions
carry both magnetic and hypermagnetic fluxes with winding 
numbers $(M, H)$, with $M$ taking odd integers $-l = -(2k+1)$
and $H=n$.
It will be shown in Sec. \ref{five} that
the hybrid vortices can be considered as
composites of the magnetic and hypermagnetic ones.

As to the topological structure of the solutions for
hybrid vortices with odd $l = 2k+1$,
a comparison of Eq. (\ref{hbtransf}) with Eq. (\ref{mtransf})
suggests the following generalization of the conclusion that
was made above for the case of the magnetic vortices: The
hybrid vortices with different $n-k$ are topologically
inequivalent.

This concludes the analysis of the topological stability of
vortex solutions.
As will be shown below, these three types of vortex solutions
exist in the present model indeed.

\section{Magnetic vortices}
\label{three}

In this section, we will analyze the solutions for the magnetic
vortices having asymptotics (\ref{masympt}) with $l$
being odd. To be concrete, we will take $l=-1$. 
In accordance with
the discussion in the previous section, one should expect that
the magnetic winding number $M$ equals +1 in this case.

Vortex solutions are static $z$-independent configurations. We
will consider the following ansatz for them in 
the cylindrical coordinates $(t, \rho, \phi, z)$
(the reasons for choosing this ansatz will
be discussed in more detail below):
\begin{eqnarray}
W^{(-)}_z(t,\rho,\phi,z) = \frac{W(\rho)e^{-i\phi}}{\sqrt{2}},
\quad  A^{(3)}_t(t,\rho,\phi,z) = A_t^{(3)}(\rho),
\quad A_{\phi}(t,\rho,\phi,z) = A_{\phi}(\rho), \nonumber\\
\quad Z_{\phi}(t,\rho,\phi,z) = Z_{\phi}(\rho),
\quad \varphi_0(t,\rho,\phi,z) = v(\rho)\,,
\label{ansatz1}
\end{eqnarray}
where $A_{\phi}$ and
$Z_{\phi}$ are angular components of the electromagnetic field
$A_{\mu}=
\frac{g^{\prime}A^{(3)}_{\mu}+gB_{\mu}}{\sqrt{g^2+g^{\prime 2}}}$ and the
$Z$-boson field
$Z_{\mu}=\frac{gA^{(3)}_{\mu}-g^{\prime}B_{\mu}}{\sqrt{g^2+g^{\prime 2}}}$
(all other components of the fields are equal to zero).
\footnote {Recall that in the cylindrical coordinates,
a vector field $\vec{A} \equiv (A_x, A_y, A_z)$ is
decomposed as $\vec{A}=A_{\rho}\hat{\rho}+A_{\phi}\hat{\phi}+
A_z\hat{z}$, 
where $A_{\rho}= A_x \cos\phi + A_y \sin\phi, \,\,
A_{\phi}= -A_x \sin\phi + A_y \cos\phi$.}

Using equations of motion (\ref{eqs-1-1})-(\ref{eqs-1-3}), one can check that
ansatz (\ref{ansatz1}) is consistent with them
and they reduce to
the following system of five coupled nonlinear ordinary differential 
equations:
\begin{equation}
W''+\frac{W'}{\rho}-\frac{W}{\rho^2}+\frac{2g(g'A_{\phi}+gZ_{\phi})}
{\sqrt{g^2+g^{'2}}}\frac{W}{\rho}+g^2\left((A_t^{(3)})^2-
\frac{(g'A_{\phi}+gZ_{\phi})^2}{g^2+g^{'2}}\right)W
-\frac{g^2}{2}Wv^2=0 \,,
\label{ode-1-1}
\end{equation}
\begin{equation}
(A_t^{(3)})''+\frac{(A_t^{(3)})'}{\rho}-g^2W^2A_t^{(3)}-
\frac{g^2}{2}A_t^{(3)} v^2+g\mu v^2=0 \,,
\label{ode-1-2}
\end{equation}
\begin{equation}
A_{\phi}''+\frac{A_{\phi}'}{\rho}-\frac{A_{\phi}}{\rho^2}+
\frac{gg'W^2}{\sqrt{g^2+g^{'2}}}
\left(\frac{1}{\rho}-\frac{g(g'A_{\phi}+gZ_{\phi})}{\sqrt{g^2+g^{'2}}}
\right)=0 \,,
\label{ode-1-3}
\end{equation}
\begin{equation}
Z_{\phi}''+\frac{Z_{\phi}'}{\rho}-\frac{Z_{\phi}}{\rho^2}-
\frac{g^2+g^{'2}}{2}Z_{\phi}v^2+
\frac{g^2W^2}{\sqrt{g^2+g^{'2}}}\left(\frac{1}{\rho}-
\frac{g(g'A_{\phi}+gZ_{\phi})}{\sqrt{g^2+g^{'2}}}\right)=0 \,,
\label{ode-1-4}
\end{equation}
\begin{equation}
v''+\frac{v'}{\rho}-\frac{g^2+g^{'2}}{4}Z_{\phi}^2v-\frac{g^2}{4}W^2v+
\left(\frac{gA_t^{(3)}}{2}-\mu\right)^2v-m^2v-2\lambda v^3=0 \,.
\label{ode-1-5}
\end{equation}
Note that although the phase in the exponent 
$e^{-i\phi}$ in $W^{(-)}_z$ in Eq.
(\ref{ansatz1}) is of the electromagnetic $U(1)_{em}$ origin, this
ansatz contains in addition 
also $Z_{\phi}$ field.
This is because the $W$ condensate mixes the gauge
fields $A_{\mu}$ and $Z_{\mu}$ (see Eqs. (\ref{ode-1-3}) and
(\ref{ode-1-4})).

It is instructive to compare ansatz (\ref{ansatz1}) and equations of motion
(\ref{ode-1-1})-(\ref{ode-1-5}) with those in the
Abelian Higgs model. Obviously, $W^{(-)}_z$ plays the role of the Higgs field.
The fields $A_{\phi}$ and $Z_{\phi}$ are analogous to the Abelian gauge field.
The fields $A_t^{(3)}$ and $v$ do
not have analogs in the Abelian Higgs model. Of course, they are present in
ansatz (\ref{ansatz1}) in order to ensure the correct
vacuum asymptotics at infinity. Thus, the fields $A_t^{(3)}$ and $v$ have a
somewhat different status in the vortex ansatz comparing to the fields 
$W^{(-)}_z$,
$A_{\phi}$, and $Z_{\phi}$. We will discuss this point in more detail 
below, when we
consider the ultraviolet boundary conditions for these fields.

The infrared
boundary conditions at $\rho \to \infty$ follow from the requirement of
finiteness of the energy per unit length 
of a vortex, which implies that
fields tend to their vacuum values:
\begin{eqnarray}
W(\rho) \to \sqrt{\frac{\sqrt{2}\mu v_0}{g} - \frac{v_0^2}{2}} \,,
\quad A_t^{(3)}(\rho) \to \frac{v_0}{\sqrt{2}} \,, \quad
v(\rho) \to v_0 \,, \quad
A_{\phi}(\rho) \to 0 \,, \quad Z_{\phi}(\rho) \to 0 \,.
\label{bc-uv1}
\end{eqnarray}

As to the ultraviolet boundary conditions at $\rho=0$,
because $\rho=0$ is a regular singular point of the system of differential
equations (\ref{ode-1-1})-(\ref{ode-1-5}), these boundary conditions  
follow from the equations themselves if one utilizes
the condition of regularity of the solution at this point:
\begin{equation}
W(0)=A_{\phi}(0)=Z_{\phi}(0)=0\,,
\label{bc-ir1-1}
\end{equation}
\begin{equation}
(A_t^{(3)})^{\prime}(0)=v^{\prime}(0)=0 \,.
\label{bc-ir1-2}
\end{equation}
The ultraviolet boundary conditions (\ref{bc-ir1-1}) for the fields $W$ 
and $A_{\phi}$, $Z_{\phi}$ are like those for a Higgs field
and a gauge field, respectively, in the
Abelian Higgs model: they guarantee that these fields are single
valued at $\rho =0$.
Unlike them, the
fields $A_t^{(3)}$ and $v$ have Neumann boundary conditions which
imply that there is no influx of momentum for these fields at the boundary
$\rho = 0$.

We solved numerically Eqs. (\ref{ode-1-1})-(\ref{ode-1-5}) with 
boundary conditions
(\ref{bc-uv1})-(\ref{bc-ir1-2}) by using {\it MATLAB}. 
To be concrete, we considered the case with $m^2 > 0$ and chose the
parameters $\mu$ and $gv_0/\sqrt 2$ as in
Ref. \cite{sigmamodel}: 
$\mu/m =1.1$ and $gv_0/\sqrt{2}m=0.1$. For the values of
the coupling constants $g$ and $g^{\prime}$ we used those from
the electroweak theory: 
$g=0.65$ and $g^{\prime}=0.35$. Then, using Eq. (\ref{v0}),
one gets the value of the coupling constant $\lambda$: $\lambda = 0.53$.
Therefore we work in the weak coupling regime when the tree
approximation is reliable.

The solution is
shown in Fig. 1, where dimensionless fields 
$W/m$, $A_t^{(3)}/m$, $A_{\phi}/m$,
$Z_{\phi}/m$, $v/m$ are plotted against $r=\rho \, m$. The
boundary conditions (\ref{bc-uv1}) were set
at $r_{max} = 80$ and it was checked that
starting from $r_{max} = 40$ the form of
the solution practically does not depend on the choice
of $r_{\max}$. 

\begin{figure}\label{fig:one}
\begin{center}
\includegraphics[scale=0.6]{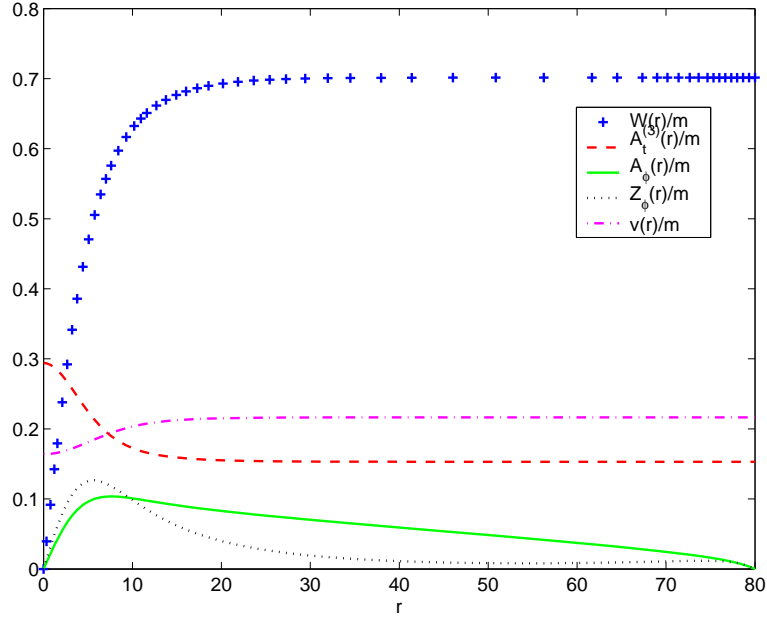}
\end{center}
\caption{Dimensionless fields $\frac{W}{m}$, $\frac{A_t^{(3)}}{m}$,
$\frac{A_{\phi}}{m}$, $\frac{Z_{\phi}}{m}$, and $\frac{v}{m}$ as functions of
$r=\rho\,m$ for the magnetic vortex solution.}
\end{figure}

Substituting ansatz (\ref{ansatz1}) into Eq. (\ref{Lagrangian}), we calculate
the effective potential ($V = - \cal{L}$):
\begin{eqnarray}
V&=&\frac{(W')^2}{2}+\frac{W^2}{2}\left(\frac{1}{\rho}-\frac{g(g'A_{\phi}+
gZ_{\phi})}{\sqrt{g^2+g^{'2}}}\right)^2-\frac{(A_t^{(3) \prime})^2}{2}
\nonumber\\
&&-\frac{g^2(A_t^{(3)})^2
W^2}{2}+\frac{1}{2}\left(A_{\phi}'+\frac{A_{\phi}}{\rho}\right)^2+
\frac{1}{2}\left(Z_{\phi}'+\frac{Z_{\phi}}{\rho}\right)^2-
\left(\frac{gA_t^{(3)}}{2}-\mu\right)^2v^2\nonumber\\
&&+(v')^2+\frac{g^2+g^{'2}}{4}Z_{\phi}^2v^2+m^2v^2+\lambda
v^4+\frac{g^2}{4}W^2v^2.
\label{V1}
\end{eqnarray}
Subtracting the energy density of the vacuum solution, we find that the energy
of the magnetic vortex per unit length is equal
to 1.27 in units of $m^2$.

Note that
the $SU(2) \times U(1)_Y$ invariant definition of the
electromagnetic field strength $F^{(em)}_{\mu\nu}$ is
\begin{equation}
F^{(em)}_{\mu\nu} = \frac{gF^{(Y)}_{\mu\nu}}{\sqrt{g^2+g^{\prime 2}}} -
\frac{2g^{\prime}}{\sqrt{g^2+g^{\prime 2}}}
\frac{\Phi^{\dagger}F^{(a)}_{\mu\nu}\frac{\tau^a}{2}\Phi}{\Phi^{\dagger}\Phi}.
\label{fieldstrength}
\end{equation}
For the vortex solution with ansatz (\ref{ansatz1}), it yields the
conventional relation for the magnetic field $\vec{H}$:
\begin{equation}
\vec{H} = \mbox{curl}\,\,\vec{A}\,.
\label{curl}
\end{equation}

It is well known that the magnetic flux of vortices is
quantized and related to the winding number,
which is 
a topological invariant. It is easy to check that this property
takes
place in the present case. Indeed, as 
follows from Eqs. (\ref{ode-1-3}) and (\ref{ode-1-4}), while
the field $A_{\phi}$ has the asymptotics $\frac{1}{e\rho}$ 
as $\rho \to \infty$ (with
$e=\frac{gg^{\prime}}{\sqrt{g^2+g^{\prime 2}}}$),
the field $Z_{\phi}$ rapidly decreases in this limit. 
Therefore we
find that for the magnetic vortex
only the flux related to the
electromagnetic field $A_{\mu}$ is nonzero. Then, using
Eq. (\ref{curl}), we get:
\begin{equation}
{\cal F}_{em}^{(M)}=\int dxdy H_{z} 
= \oint d\vec{x} \cdot  \vec{A} = \int d\phi \,\,
\rho A_{\phi}|_{\rho=\infty} = \frac{2\pi}{e}.
\label{flux1}
\end{equation}
I.e., as was expected, this vortex solution corresponds to
the winding number +1.

\section{Hypermagnetic vortices}
\label{four}

A hypermagnetic vortex possesses asymptotics (\ref{hasympt}).
To be concrete, we will take $n=-1$.
As follows from the discussion
in Sec. \ref{two}, the hypermagnetic winding number $H$ should be
now equal to -1.

In this case, we will use the following ansatz:
\begin{eqnarray}
W^{(-)}_z(t,\rho,\phi,z) = \frac{W(\rho)}{\sqrt{2}},
\quad  A^{(3)}_t(t,\rho,\phi,z) = A_t^{(3)}(\rho),
\quad A_{\phi}(t,\rho,\phi,z) = A_{\phi}(\rho), \nonumber\\
\quad Z_{\phi}(t,\rho,\phi,z) = Z_{\phi}(\rho),
\quad \varphi_0(t,\rho,\phi,z) = v(\rho)e^{-i\phi}\,.
\label{ansatz2}
\end{eqnarray}
For this ansatz, the equations of motion (\ref{eqs-1-1})-(\ref{eqs-1-3}) yield
\begin{equation}
W''+\frac{W'}{\rho}+g^2\left((A_t^{(3)})^2-\frac{(g'A_{\phi}+gZ_{\phi})^2}
{g^2+g^{'2}}\right)W-\frac{g^2}{2}Wv^2=0 \,,
\label{ode2-1}
\end{equation}
\begin{equation}
(A_t^{(3)})''+\frac{(A_t^{(3)})'}{\rho}-g^2W^2A_t^{(3)}-
\frac{g^2}{2}A_t^{(3)} v^2+g\mu v^2=0 \,,
\label{ode2-2}
\end{equation}
\begin{equation}
A_{\phi}''+\frac{A_{\phi}'}{\rho}-\frac{A_{\phi}}{\rho^2}-
\frac{g^2g'(g'A_{\phi}+gZ_{\phi})}{g^2+g^{'2}}W^2=0 \,,
\label{ode2-3}
\end{equation}
\begin{equation}
Z_{\phi}''+\frac{Z_{\phi}'}{\rho}-\frac{Z_{\phi}}{\rho^2}+
\frac{\sqrt{g^2+g^{'2}}}{2}\left(\frac{2}{\rho}-
\sqrt{g^2+g^{'2}}Z_{\phi}\right)v^2-\frac{g^3(g'A_{\phi}+gZ_{\phi})}
{g^2+g^{'2}}W^2=0 \,,
\label{ode2-4}
\end{equation}
\begin{equation}
v''+\frac{v'}{\rho}-\frac{1}{4}\left(\frac{2}{\rho}-
\sqrt{g^2+g^{'2}}Z_{\phi}\right)^2v-\frac{g^2}{4}W^2v+
\left(\frac{gA_t^{(3)}}{2}-\mu\right)^2v-m^2v-2\lambda
v^3=0.
\label{ode2-5}
\end{equation}

The infrared
boundary conditions for the hypermagnetic vortex ansatz at $\rho \to \infty$
are the same as for the magnetic vortex and given in 
Eq. (\ref{bc-uv1}).
The ultraviolet boundary conditions at $\rho = 0$ are
\begin{equation}
v(0)=A_{\phi}(0)=Z_{\phi}(0)=0\,,
\label{bc-ir2-1}
\end{equation}
\begin{equation}
(A_t^{(3)})^{\prime}(0)=W^{\prime}(0)=0 \,.
\label{bc-ir2-2}
\end{equation}
Comparing these boundary conditions with those
in Eqs. (\ref{bc-ir1-1}) and (\ref{bc-ir1-2})
for the magnetic vortex, one can see that
the ultraviolet boundary conditions for $W$ and $v$ are
now interchanged. Of course, this point is
related to the fact that the phase in the scalar field 
$\varphi_0$ (and not vector field $W$) is relevant
for the hypermagnetic $U(1)_Y$ vortex. The
numerical solution of
Eqs. (\ref{ode2-1})-(\ref{ode2-5}) with boundary conditions (\ref{bc-uv1}),
(\ref{bc-ir2-1}), and (\ref{bc-ir2-2}) is shown in Fig. 2, where we use the
same set of parameters as for the magnetic vortex.

\begin{figure}\label{fig:two}
\begin{center}
\includegraphics[scale=0.6]{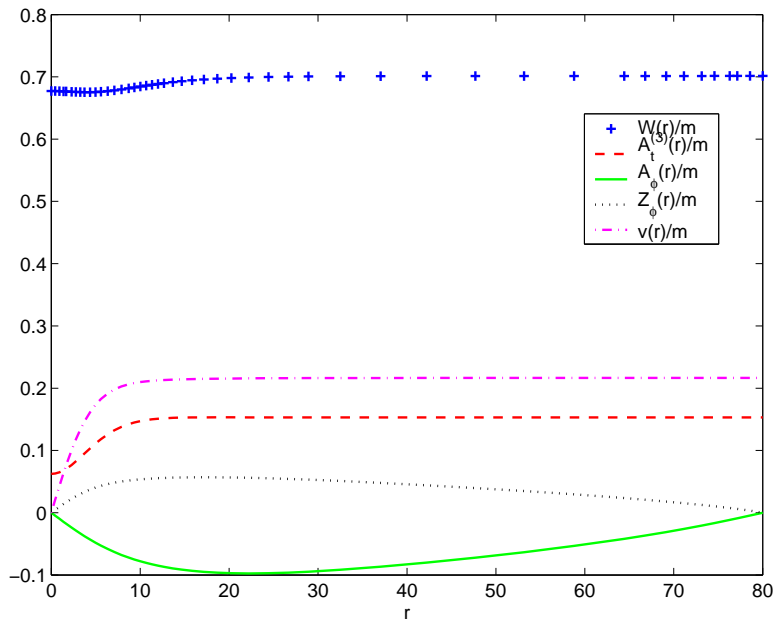}
\end{center}
\caption{Dimensionless fields $\frac{W}{m}$, $\frac{A_t^{(3)}}{m}$,
$\frac{A_{\phi}}{m}$, $\frac{Z_{\phi}}{m}$, and $\frac{v}{m}$ as functions of
$r=\rho\,m$ for the hypermagnetic vortex solution.}
\end{figure}

The effective potential for ansatz (\ref{ansatz2}) is
\begin{eqnarray}
V&=&\frac{(W')^2}{2}+\frac{g^2}{2}\frac{(g'A_{\phi}+gZ_{\phi})^2}
{g^2+g^{'2}}W^2-\frac{(A_t^{(3) '})^2}{2}\nonumber\\
&&-\frac{g^2(A_t^{(3)})^2
W^2}{2}+\frac{1}{2}\left(A_{\phi}'+\frac{A_{\phi}}{\rho}\right)^2+
\frac{1}{2}\left(Z_{\phi}'+\frac{Z_{\phi}}{\rho}\right)^2-
\left(\frac{gA_t^{(3)}}{2}-\mu\right)^2v^2\nonumber\\
&&+(v')^2+\frac{1}{4}\left(\frac{2}{\rho}-
\sqrt{g^2+g^{'2}}Z_{\phi}\right)^2v^2+m^2v^2+\lambda
v^4+\frac{g^2}{4}W^2v^2 \,.
\end{eqnarray}
Subtracting the energy density of the vacuum solution, we find that the
energy density of the 
hypermagnetic vortex per unit length in units of $m^2$ is
equal to 0.79.

It follows from Eqs. (\ref{ode2-3}) and (\ref{ode2-4}) that $A_{\phi}$ and
$Z_{\phi}$ have the asymptotics
$-\frac{2g}{g^{\prime}\rho\sqrt{g^2+g^{\prime 2}}}$ and
$\frac{2}{\rho\sqrt{g^2+g^{\prime 2}}}$ as $\rho \to \infty$. Therefore,
the fluxes of both fields $A_{\phi}$ and $Z_{\phi}$ are nonzero
\begin{equation}
{\cal F}_{em}^{(H)} = -\frac{4 \pi g}{g^{\prime}\sqrt{g^2+g^{\prime 2}}} \,,
\label{flux2-em}
\end{equation}
\begin{equation}
{\cal F}_{Z}^{(H)} = \frac{4 \pi}{\sqrt{g^2+g^{\prime 2}}} \,.
\label{flux2-Z}
\end{equation}
It is not difficult to see that these two fluxes correspond to the
hypermagnetic flux of the 
$U(1)_Y$ gauge field $B_{\mu}$ (the hypermagnetic flux of the
$A^{(3)}_{\mu}$ field is zero):
\begin{equation}
{\cal F}_{Y}^{(H)} = -\frac{4 \pi}{g^{\prime}} \,.
\label{flux2-Y}
\end{equation}
Note that
because the coupling of the Higgs field with the field
$B_{\mu}$ in the covariant derivative
$D_{\mu}=\partial_{\mu}-igA_{\mu}-(ig^{\prime}/2)B_{\mu}$ is 
expressed through
$g^{\prime}/2$ rather than $g^{\prime}$, this flux corresponds to
the winding number -1, as was expected.

\section{Hybrid vortices}
\label{five}

The hybrid vortex solutions carry both the magnetic and hypermagnetic
fluxes with winding numbers $(M, H)$. Their asymptotics are given
in Eq. (\ref{hbasympt}). To be concrete, we will consider the
solutions with $l=-1, n=\mp 1$ corresponding to
$(M, H)=(1, \mp 1)$, i.e., we consider 
hybrid vortices with the
equal and opposite phases of the $W^{(-)}_z$ and $\varphi_0$ fields.

Clearly, the ansatz for these vortices should have the form
\begin{eqnarray}
W^{(-)}_z(t,\rho,\phi,z) = \frac{W(\rho)e^{-i\phi}}{\sqrt{2}},
\quad  A^{(3)}_t(t,\rho,\phi,z) = A_t^{(3)}(\rho),
\quad A_{\phi}(t,\rho,\phi,z) = A_{\phi}(\rho), \nonumber\\
\quad Z_{\phi}(t,\rho,\phi,z) = Z_{\phi}(\rho),
\quad \varphi_0(t,\rho,\phi,z) = v(\rho)e^{-i\sigma\phi}\,,
\label{ansatz3}
\end{eqnarray}
where $\sigma=\pm$.

The equations of motion for these vortices are
sort of a combination of the equations of motion for the magnetic and
hypermagnetic ones:
\begin{equation}
W''+\frac{W'}{\rho}-\frac{W}{\rho^2}+\frac{2g(g'A_{\phi}+gZ_{\phi})}
{\sqrt{g^2+g^{'2}}}\frac{W}{\rho}+g^2\left((A_t^{(3)})^2-
\frac{(g'A_{\phi}+gZ_{\phi})^2}{g^2+g^{'2}}\right)W
-\frac{g^2}{2}Wv^2=0 \,,
\label{ode3-1}
\end{equation}
\begin{equation}
(A_t^{(3)})''+\frac{(A_t^{(3)})'}{\rho}-g^2W^2A_t^{(3)}-
\frac{g^2}{2}A_t^{(3)} v^2+g\mu v^2=0 \,,
\label{ode3-2}
\end{equation}
\begin{equation}
A_{\phi}''+\frac{A_{\phi}'}{\rho}-\frac{A_{\phi}}{\rho^2}+
\frac{gg'W^2}{\sqrt{g^2+g^{'2}}}
\left(\frac{1}{\rho}-\frac{g(g'A_{\phi}+gZ_{\phi})}{\sqrt{g^2+g^{'2}}}
\right)=0 \,,
\label{ode3-3}
\end{equation}
\begin{equation}
Z_{\phi}''+\frac{Z_{\phi}'}{\rho}-\frac{Z_{\phi}}{\rho^2}+
\frac{\sqrt{g^2+g^{'2}}}{2}(\frac{2\sigma}{\rho}-\sqrt{g^2+g^{'2}}Z_{\phi})v^2+
\frac{g^2W^2}{\sqrt{g^2+g^{'2}}}\left(\frac{1}{\rho}-
\frac{g(g'A_{\phi}+gZ_{\phi})}{\sqrt{g^2+g^{'2}}}\right)=0 \,,
\label{ode3-4}
\end{equation}
\begin{equation}
v''+\frac{v'}{\rho}-\frac{(\frac{2\sigma}{\rho}-
\sqrt{g^2+g^{'2}}Z_{\phi})^2v}{4}-\frac{g^2}{4}W^2v+
\left(\frac{gA_t^{(3)}}{2}-\mu\right)^2v-m^2v-2\lambda v^3=0 \,.
\label{ode3-5}
\end{equation}

The infrared boundary conditions at infinity remain the same
as before.
Requiring the regularity of solutions at $\rho=0$
leads to the following ultraviolet boundary conditions:
\begin{equation}
W(0)=v(0)=A_{\phi}(0)=Z_{\phi}(0)=0\,,
\label{bc-ir3-1}
\end{equation}
\begin{equation}
(A_t^{(3)})^{\prime}(0) = 0 \,.
\label{bc-ir3-2}
\end{equation}
Note that in contrast to the cases of the magnetic and 
hypermagnetic vortices,
now both functions $W(\rho)$ and $v(\rho)$ have  
the Dirichlet boundary
conditions at $\rho=0$. The reason of this is
clear: Because both $W^{(-)}_z$ and
$\varphi_0$ have nontrivial phases for hybrid vortices (\ref{ansatz3}), the
regularity of solutions at $\rho=0$ can be ensured only if the moduli of these
fields vanish at this point.

The numerical solutions of
Eqs.(\ref{ode3-1})-(\ref{ode3-5}) with boundary conditions (\ref{bc-uv1}),
(\ref{bc-ir3-1}), and (\ref{bc-ir3-2}) are shown in Fig. 3 and Fig. 4. 
For these solutions, the same set of parameters was used as
in the previous cases.

\begin{figure}\label{fig:three}
\begin{center}
\includegraphics[scale=0.6]{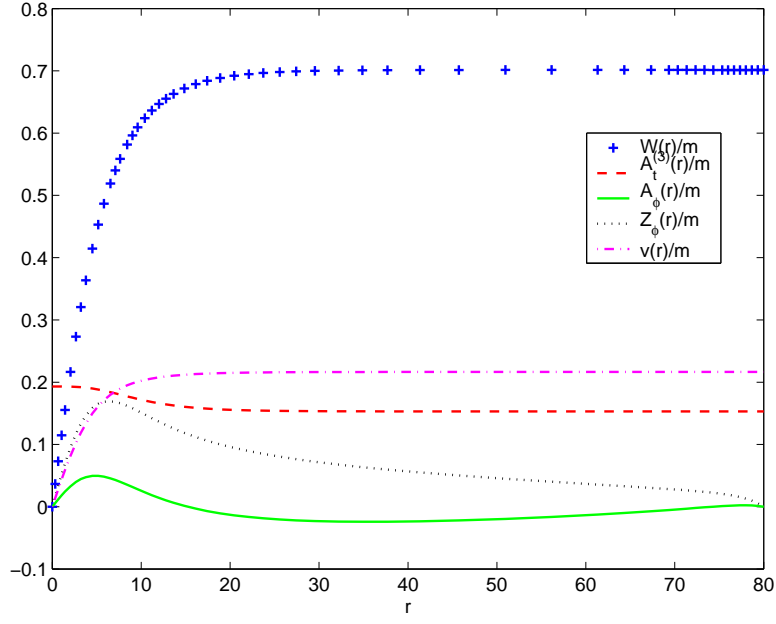}
\end{center}
\caption{Dimensionless fields $\frac{W}{m}$, $\frac{A_t^{(3)}}{m}$,
$\frac{A_{\phi}}{m}$, $\frac{Z_{\phi}}{m}$, and $\frac{v}{m}$ as functions of
$r=\rho\,m$ for the hybrid vortex solution with equal phases.}
\end{figure}

\begin{figure}\label{fig:four}
\begin{center}
\includegraphics[scale=0.6]{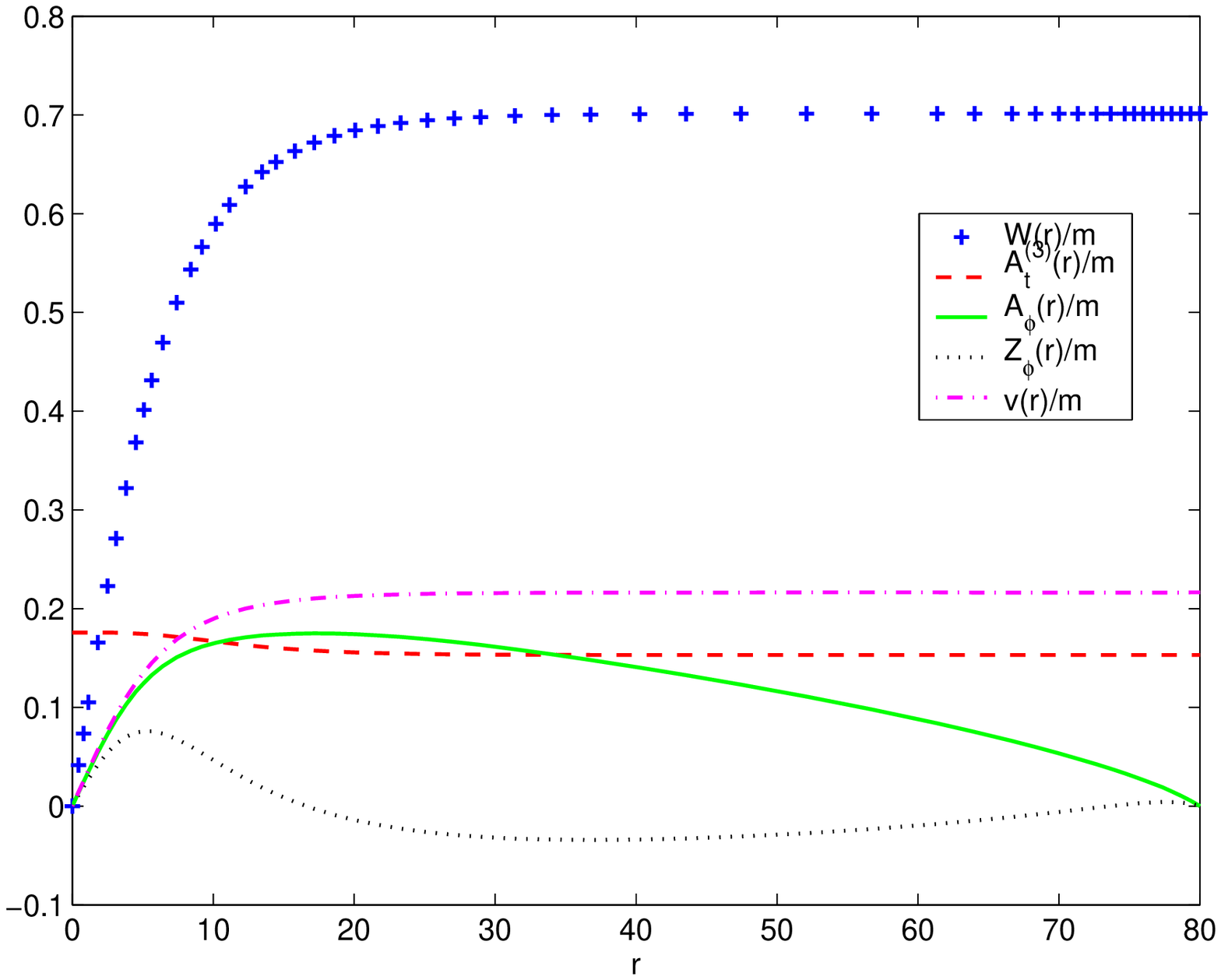}
\end{center}
\caption{Dimensionless fields $\frac{W}{m}$, $\frac{A_t^{(3)}}{m}$,
$\frac{A_{\phi}}{m}$, $\frac{Z_{\phi}}{m}$, and $\frac{v}{m}$ as functions of
$r=\rho\,m$ for the hybrid vortex solution with opposite phases.}
\end{figure}

The effective potential for ansatz (\ref{ansatz3}) is equal to
\begin{eqnarray}
V&=&\frac{(W')^2}{2}+\frac{W^2}{2}(\frac{1}{\rho}-
\frac{g(g'A_{\phi}+gZ_{\phi})}{\sqrt{g^2+g^{'2}}})^2
-\frac{(A_t^{(3)'})^2}{2}\nonumber\\
&&-\frac{g^2(A_t^{(3)})^2
W^2}{2}+\frac{1}{2}\left(A_{\phi}'+\frac{A_{\phi}}{\rho}\right)^2+
\frac{1}{2}\left(Z_{\phi}'+\frac{Z_{\phi}}{\rho}\right)^2-
\left(\frac{gA_t^{(3)}}{2}-\mu\right)^2v^2\nonumber\\
&&+(v')^2+\frac{1}{4}\left(\frac{2\sigma}{\rho}-
\sqrt{g^2+g^{'2}}Z_{\phi}\right)^2v^2+m^2v^2+\lambda
v^4+\frac{g^2}{4}W^2v^2 \,.
\end{eqnarray}
Subtracting the energy density of the vacuum solution, we find that 
in units of $m^2$, the
energy per unit length of hybrid vortices with the equal and
opposite phases are equal to 1.49 and 2.3, respectively.

It follows from Eqs. (\ref{ode3-3}) and (\ref{ode3-4}) that $A_{\phi}$ and
$Z_{\phi}$ have the asymptotics
$\frac{g^2(1-2\sigma)+g^{\prime 2}}{\rho gg^{\prime}\sqrt{g^2+g^{\prime 2}}}$ 
and
$\frac{2\sigma}{\rho\sqrt{g^2+g^{\prime 2}}}$ as $\rho \to \infty$. Therefore,
the fluxes of hybrid vortices are equal to
\begin{equation}
{\cal F}_{em}^{(MH)} = \frac{2\pi[g^2(1-2\sigma)+g^{\prime 2}]}
{gg^{\prime}\sqrt{g^2+g^{\prime 2}}} \,,
\label{flux-hybrid-em}
\end{equation}
\begin{equation}
{\cal F}_{Z}^{(MH)} = \frac{4 \pi \sigma}{\sqrt{g^2+g^{\prime 2}}} \,.
\label{flux-hybrid-Z}
\end{equation}

These expressions can be 
rewritten in the following transparent way:
\begin{equation}
{\cal F}_{em}^{(MH)} = 1\,\cdot {\cal F}_{em}^{(M)}
+ \sigma \cdot {\cal F}_{em}^{(H)}\,,
\label{decomposition1}
\end{equation}
\begin{equation}
{\cal F}_{Z}^{(MH)} = 1\,\cdot {\cal F}_{Z}^{(M)}
+ \sigma \cdot {\cal F}_{Z}^{(H)}\,,
\label{decomposition2}
\end{equation}
where ${\cal F}_{em,Z}^{(M)}$ and ${\cal F}_{em,Z}^{(H)}$ are the fluxes
of the magnetic and hypermagnetic vortices given by Eqs.(\ref{flux1}) and
(\ref{flux2-em}), (\ref{flux2-Z}), respectively (recall that
${\cal F}_{Z}^{(M)}=0$). This suggests that 
the hybrid vortices can be considered as
composites of the magnetic and hypermagnetic ones. 
Note that the sum of the energies per unit length for
the magnetic vortex and the hypermagnetic one is equal to
$1.27 + 0.79= 2.06$. On the other hand,
because the energies for the hybrid
vortices with $\sigma =+$ and $\sigma =-$ are 
equal to 1.49 and 2.3, respectively,  
they satisfy the inequality $1.49 < 2.06 < 2.3$. This inequality
implies that
while the hybrid vortex with equal phases is stable, the hybrid
vortex with opposite phases is not.

\section{Conclusion}
\label{six}

The set of vortex solutions we obtained   
in the gauged $SU(2)\times U(1)_Y$
$\sigma$-model with the chemical potential for hypercharge 
is quite rich. 
In particular, there are different types of
vortices connected either with photon field or hypercharge
gauge field, or with both of them.
The richness of this set
is provided by the structure of the ground state (\ref{vacuum}),
which includes vector condensates. This ground state describes
an anisotropic medium with electric superconductivity. 

It is quite noticeable that the sample of symmetry breaking in this
model is the same as in the gluonic phase in neutral two-flavor QCD
\cite{gluonic}, while the present model is much simpler. 
The point is that because
of a nonzero baryon density in the gluonic phase, its dynamics
in the hard-dense-loop approximation
is mostly provided by quark loops. And since most of the initial
symmetries (including rotational $SO(3)_{rot}$) in this phase are
spontaneously broken, the calculation of its effective action
is quite involved. This makes the study of vortex solutions
in the gluonic phase to be quite a difficult problem. 
On the other hand, vortices could be very important
for understanding the dynamics of quark matter in compact stars.
The existence of such solutions in a related but much simpler model
is encouraging.  

It is instructive to compare the present vortex solutions with
those in such complex condensed matter systems as the 
$B$ phase in $^{3}\mbox{He}$ \cite{Vol} and
high $T_c$ superconductors \cite{Sachd}. Unlike the conventional 
Abrikosov-Nielsen-Olesen vortices \cite{Abrikosov,NO}, the initial
symmetries are not completely restored in the core of vortices in those 
systems. In $^{3}\mbox{He}$, vortices with an A-phase core or
with asymmetric core are realized. In high $T_c$ superconductors,
there is an antiferromagnetic condensate in the vortex core. A 
similar situation takes place for the magnetic and hypermagnetic 
vortices in the present model. In the core of a magnetic vortex, while
the order parameter $W^{(-)}_z$ is zero,
the order parameter $\varphi_0$ is not (see Fig. 1). Therefore, in its core
only the $U(1)_{em}$ symmetry is restored. In the core 
of a hypermagnetic vortex, we found that while $\varphi_0 = 0$,
the order parameter $W^{(-)}_z$ is not zero (see Fig. 2). 
Therefore, only the
$U(1)_{Y}$ symmetry is restored in the core in this case. 
On the other hand, the whole abelian  
$U(1)_{em}\times U(1)_{Y}$ gauge symmetry
is restored in the core of a hybrid vortex (see Figs. 3 and 4). 
 
Recently,
there has been a considerable interest in systems with coexisting
order parameters (such as
high $T_c$ superconductors) in condensed matter \cite{Sachd}. 
Generating vector condensates is a very natural way of creating
such systems (for example, in the present model, 
electric superconductivity
coexists with spontaneous rotational symmetry breaking). 
This possibility deserves further study.  

\section{Acknowledgements}

One of us (V.A.M.) acknowledges an inspiring question of D. T. Son
concerning a possibility of the existence of
vortex solutions in the linear $\sigma$ model
with the chemical potential for hypercharge. 
We thank V. P. Gusynin, M. Hashimoto, and
I. A. Shovkovy for useful discussions. 
This work was supported by the Natural Sciences and Engineering Research
Council of Canada.

\end{document}